\documentclass[conference,a4paper]{IEEEtran}

\makeatletter
 \def\LaTeX{\leavevmode L\raise.42ex
   \hbox{\kern-.3em\size{\sf@size}{0pt}\selectfont A}\kern-.15em\TeX}
\makeatother

\newcommand{\BibTeX}{{\rm B\kern-.05em{\sc
i\kern-.025emb}\kern-.08em\TeX}}

\newtheorem{thm}{Theorem}[section]
\newtheorem{lem}[thm]{Lemma}
\newtheorem{rem}{Remark}
\newtheorem{defn}{Definition}[section]

\begin{document}

\title{Shannon Sampling and Parseval \\ Frames on Compact  Manifolds}

\author{\IEEEauthorblockN{Isaac Z. Pesenson}
\IEEEauthorblockA{
 Temple University  \\
Philadelphia, USA\\
Email: pesenson@temple.edu}}

\maketitle


\IEEEpeerreviewmaketitle

\section{Introduction}

	The problem of representation and analysis of functions defined on manifolds (signals, images, and data in general) is ubiquities in many fields ranging from statistics and cosmology to neuroscience and  biology.  It is  very common  to consider  input signals as points in a high-dimensional measurement space, however,  meaningful structures lay on a manifold embedded in this space. 	
	
		In the last decades, the importance of these applications triggered the development of various generalized wavelet bases suitable for the unit spheres $S^{2}$ and $S^{3}$ and  the rotation group of $\mathbf{R}^{3}$. The goal of the present study is to describe a general approach  to  bandlimited  localized Parseval frames in a space $L_{2}(\mathbf{\mathbf{M}})$, where $\mathbf{\mathbf{M}}$ is a compact homogeneous Riemannian manifold.

	One can think of a Riemannian manifold as of a surface in a Euclidean space. A homogeneous manifold is a surface with "many" symmetries like the sphere $x_{1}^{2}+...+x_{d}^{2}=1$ in Euclidean space $\mathbf{R}^{d}$. 
	
	Our construction   of frames in a function space $L_{2}(\mathbf{\mathbf{M}})$ heavily depends on proper notions of bandlimitedness and Shannon-type sampling on a manifold $\mathbf{\mathbf{M}}$. 
	The crucial role in this development is played by positive cubature formulas (Theorem \ref{cubature}) and by the product property (Theorem \ref{product}), which were proved in \cite{gpes} and \cite{pesg}.

The notion of bandlimideness on a compact manifold $\mathbf{\mathbf{M}}$ is introduced in terms of eigenfunctions of a certain second-order differential elliptic operator on $\mathbf{\mathbf{M}}$. 
 The most  important fact for our construction of frames  is that in a space of $\omega$-bandlimited functions 
the regular $L_{2}(\mathbf{M})$ norm can be descretized. This result in the case of compact manifolds (and even non-compact manifolds of bounded geometry) was first discovered and explored in many ways in our papers \cite{Pes98}-\cite{Pes09}. In the classical cases of straight line $\mathbf{R}$ and  circle $\mathbf{S}$ the corresponding results are known as Plancherel-Polya and  Marcinkiewicz-Zygmund inequalities. Our generalization of Plancherel-Polya and  Marcinkiewicz-Zygmund inequalities  implies  that $\omega$-bandlimited functions  on manifolds are completely determined by their vales on discrete sets of points "uniformly" distributed over $\mathbf{M}$ with a spacing comparable to $1/\sqrt{\omega}$ and can be completely reconstructed in a stable way from their values on such sets. The last statement is an extension of the  Shannon sampling theorem to the case of Riemannian manifolds.

  Our article is a summary of some results for Riemannian  manifolds that were obtained in   \cite{gpes}-\cite{Pesssubm}.   To the best of our knowledge these are the pioneering papers which contain the most general   results about frames,  Shannon sampling, and cubature formulas on compact and non-compact Riemannian manifolds. In particular, the paper \cite{gpes} gives an "end point" construction of tight localized frames on homogeneous compact manifolds.  The paper \cite{Pessubm} is the first systematic development of localized frames on compact domains in Euclidean spaces.

\subsection { Compact homogeneous manifolds}

A homogeneous compact manifold $\mathbf{\mathbf{M}}$ is a
$C^{\infty}$-compact manifold on which a compact
Lie group $G$ acts transitively. In this case $\mathbf{\mathbf{M}}$ is necessary of the form $G/H$,
where $H$ is a closed subgroup of $G$. The notation $L_{2}(\mathbf{\mathbf{M}}),$ is used for the usual Hilbert spaces, where $dx$ is an invariant
measure.

If $\textbf{g}$ is the Lie algebra of a compact Lie group $G$ then  it is a direct sum
$\textbf{g}=\textbf{a}+[\textbf{g},\textbf{g}]$, where
$\textbf{a}$ is the center of $\textbf{g}$, and
$[\textbf{g},\textbf{g}]$ is a semi-simple algebra. Let $Q$ be a
positive-definite quadratic form on $\textbf{g}$ which, on
$[\textbf{g},\textbf{g}]$, is opposite to the Killing form. Let
$X_{1},...,X_{d}$ be a basis of
$\textbf{g}$, which is orthonormal with respect to $Q$.
 Since the form $Q$ is $Ad(G)$-invariant, the operator
$$
-X_{1}^{2}-X_{2}^{2}-\    ... -X_{d}^{2},    \ d=dim\ G
$$
is a bi-invariant operator on $G$, which is known as the Casimir operator. This implies in particular that
the
   corresponding operator on $L_{2}(\mathbf{\mathbf{M}}), $
\begin{equation}\label{Casimir}
L=-D_{1}^{2}- D_{2}^{2}- ...- D_{d}^{2}, \>\>\>
       D_{j}=D_{X_{j}}, \        d=dim \ G,
\end{equation}
commutes with all operators $D_{j}=D_{X_{j}}$.
Operator $L$, which is usually called the Laplace operator, is
the image of the Casimir operator under differential of quasi-regular representation in $L_{2}(\mathbf{\mathbf{M}})$.
Note that if $\mathbf{\mathbf{M}}=G/H$ is a compact symmetric space then the number
$d=dim\> G$ of operators in the formula (\ref{Casimir}) can be
strictly bigger than the dimension $ n=dim\> \mathbf{\mathbf{M}}$. For example on a
two-dimensional sphere $\mathbf{S}^{2}$ the Laplace-Beltrami
operator $L_{\mathbf{S}^{2}}$ is written as
$
L_{\mathbf{S}^{2}}=D_{1}^{2}+ D_{2}^{2}+
D_{3}^{2},
$
where $D_{i}, i=1,2,3,$ generates a rotation  in $\mathbf{R}^{3}$
around coordinate axis $x_{i}$:
$
D_{i}=x_{j}\partial_{k}-x_{k}\partial_{j},
$
where $j,k\neq i.$

It is important to realize that in general, the operator $L$ is not necessarily the Laplace-Beltrami operator of the natural  invariant metric on $\mathbf{\mathbf{M}}$. But it coincides with such operator at least in the following cases:
\begin{enumerate}

\item  If the manifold $\mathbf{\mathbf{M}}$ is itself a compact Lie group $G$  then $L$ is exactly the Laplace-Beltrami
operator of an invariant metric on $G$. In particular it happens if $\mathbf{\mathbf{M}}$ is an $n$-dimensional torus, and $L$ is the sum
of squares of partial derivatives;

\item  If $\mathbf{\mathbf{M}}=G/H$ is a compact symmetric space of
  rank one, then the operator
 $L$ is proportional to the Laplace-Beltrami operator
of an invariant metric on $G/H$. This follows from the fact that, in
the rank one case, every second-order operator  which commutes with
all isometries $x\rightarrow g\cdot x, \>\>\>x\in  \mathbf{\mathbf{M}},\>\>\> g\in
G,$ is proportional to the Laplace-Beltrami operator. The important examples of such manifolds are spheres and  projective spaces.
\end{enumerate}

Since manifold $\mathbf{\mathbf{M}}$ is compact and $L$ is  a second-order differential   
elliptic self-adjoint positive definite operator $L_{2}(\mathbf{\mathbf{M}})$   it has a discrete spectrum $0=\lambda_{0}<\lambda_{1}\leq \lambda_{2}\leq......$ which goes to infinity and there exists a complete  family  $\{u_{j}\}$  of orthonormal eigenfunctions which form a  basis in $L_{2}(\mathbf{\mathbf{M}})$. 

 \begin{defn}
The span of eigenfunctions $u_{j}$ 
 $$
 Lu_{j}=\lambda_{j}u_{j}
 $$
 with $\lambda_{j}\leq \omega,\>\>\>\omega>0,$ is denoted as  $\mathbf{\mathbf{E}}_{\omega}(L)$ and is called the space of bandlimited functions on $\mathbf{\mathbf{M}}$ of bandwidth $\omega$.
 \end{defn}

According to the  Weyl's asymptotic formula one has
\begin{equation}\label{Weyl}
dim \>\mathbf{\mathbf{E}}_{\omega}(L)\sim C \>Vol(\mathbf{\mathbf{M}})\omega^{n/2},
\end{equation}
where $n=dim \>\mathbf{\mathbf{M}}$ and $C$ is an absolute constant.

Let $B(x,r)$ be a metric ball on a compact Riemannian manifold $\mathbf{\mathbf{M}}$ whose center is $x$ and
radius is $r$. The following lemma can be found  in  \cite{Pes00},
\cite{Pes04b}.

\begin{lem}
There exists
a natural number $N_{\mathbf{M}}$, such that  for any sufficiently small $\rho>0$,
there exists a set of points $\{x_{k}\}$ such that:
\begin{enumerate}
\item the balls $B(x_{k}, \rho/4)$ are disjoint,

\item  the balls $B(x_{k}, \rho/2)$ form a cover of $\mathbf{M}$,

\item  the multiplicity of the cover by balls $B(x_{k}, \rho)$
is not greater than $N_{\mathbf{M}}.$
\end{enumerate}\label{covL}
\end{lem}

\begin{defn}
Any set of points $\mathbf{M}_{\rho}=\{x_{k}\}$ which is  described in
Lemma \ref{covL} will be called a metric
$\rho$-lattice.\label{D1}
\end{defn}

The following theorems are of primary  importance.

\begin{thm}\label{product}(Product property \cite{gpes}, \cite{pesg})
\label{prodthm}
If $\mathbf{M}=G/H$ is a compact homogeneous manifold and $L$
 is the same as above, then for any $f$ and $g$ belonging
to $\mathbf{E}_{\omega}(L)$,  their product $fg$ belongs to
$\mathbf{E}_{4d\omega}(L)$, where $d$ is the dimension of the
group $G$.

\begin{rem}
At this moment it is not known if the constant $4d$ can be lowered in general situation.  However, it is easy to verify  that in the case of two-point homogeneous manifolds (which include spheres and projective spaces) a stronger result holds: if $f, \>g\in \mathbf{E}_{\omega}(L)$ then $fg\in \mathbf{E}_{2\omega}(L)$.

\end{rem}

\end{thm}
\begin{thm}\label{cubature} (Cubature  formula \cite{gpes}, \cite{pesg})
There exists  a  positive constant $c=c(\mathbf{M})$,    such  that if  $\rho=c\omega^{-1/2}$, then
for any $\rho$-lattice $\mathbf{M}_{\rho}$, there exist strictly positive coefficients $\alpha_{x_{k}}>0, 
 \  x_{k}\in \mathbf{\mathbf{M}}_{\rho}$, \  for which the following equality holds for all functions in $ \mathbf{\mathbf{E}}_{\omega}(\mathbf{M})$:
\begin{equation}
\label{cubway}
\int_{\mathbf{\mathbf{M}}}fdx=\sum_{x_{k}\in \mathbf{\mathbf{M}}_{\rho}}\alpha_{x_{k}}f(x_{k}).
\end{equation}
Moreover, there exists constants  $\  c_{1}, \  c_{2}, $  such that  the following inequalities hold:
\begin{equation}
c_{1}\rho^{n}\leq \alpha_{x_{k}}\leq c_{2}\rho^{n}, \ n=dim\ \mathbf{M}.
\end{equation}
\end{thm}

\section{Hilbert frames}

Since eigenfunctions have perfect localization properties in the spectral domain they cannot be localized on the manifold. 

It is the goal of our development to construct "better bases" in corresponding $L_{2}(\mathbf{M})$ spaces which will have rather strong localization on a manifold and  in the spectral domain.

In fact, the "kind of basis" which we are going to construct   is known today as a frame.

A set of vectors $\{\psi_{v}\}$  in a Hilbert space $\mathcal{H}$ is called a frame if there exist constants $A, B>0$ such that for all $f\in \mathcal{H}$ 
\begin{equation}
A\|f\|^{2}_{2}\leq \sum_{v}\left|\left<f,\psi_{v}\right>\right|^{2}     \leq B\|f\|_{2}^{2}.
\end{equation}
The largest $A$ and smallest $B$ are called lower and upper frame bounds.

The set of scalars $\{\left<f,\psi_{v}\right>\}$ represents a set of measurements of a signal $f$. To synthesize signal $f$ from this set of measurements one has to find another (dual) frame $\{\Psi_{v}\}$ and then a reconstruction formula is 
\begin{equation}\label{Hfr}
f=\sum_{v}\left<f,\psi_{v}\right>\Psi_{v}.
\end{equation}

Dual frame is not unique in general.  Moreover it is difficult   to find a dual frame.
If in particular $A=B=1$ the frame is said to be  tight   or Parseval. 

The main feature of Parseval frames is that 
decomposing
and synthesizing a signal or image from known data are tasks carried out with
the same set of functions. In other words in (\ref{Hfr}) one can have $\Psi_{v}=\psi_{\nu}$.

Parseval frames are similar in many respects to orthonormal wavelet bases.  For example, if in addition all vectors $\psi_{v}$ are unit vectors, then the frame is an  orthonormal basis. 
However, the important differences between frames and, say, orthonormal bases is their  redundancy that helps reduce the effect of noise in data.

Frames in Hilbert spaces of functions  whose members have simultaneous localization in space and frequency  arise naturally in wavelet analysis on  Euclidean spaces  when continuous wavelet transforms are discretized.
 Such frames have been constructed, studied, and
employed extensively in both theoretical and applied problems.

\section{Bandlimited localized Parseval frames on compact homogeneous manifolds}

According to spectral theorem if $F$ is a Schwartz function on the line, then there is a well defined operator $F(L)$ in the space $L_{2}(\mathbf{\mathbf{M}})$ such that for any $f\in L_{2}(\mathbf{\mathbf{M}})$ one has 
\begin{equation}\label{function}
\left(F(L)f\right)(x)=\int_{\mathbf{\mathbf{M}}}\mathcal{K}^{F}(x,y)f(y)dy,
\end{equation}
where $dy$ is the invariant normalized measure on $\mathbf{\mathbf{M}}$ and 
\begin{equation}\label{kernel-0}
\mathcal{K}^{F}(x,y)=\sum_{j=0}^{\infty}F(\lambda_{j})u_{j}(x)\overline{u_{j}}(y).
\end{equation}
We will be especially interested in operators of the form $F(t^{2}L)$, where $F$ is a Schwartz function and $t>0$. The corresponding kernel will be denoted as $\mathcal{K}_{t}^{F}(x,y)$ and 
\begin{equation}\label{kernel-1}
\mathcal{K}_{t}^{F}(x,y)=\sum_{j=0}^{\infty}F(t^{2}\lambda_{j})u_{j}(x)\overline{u_{j}}(y).
\end{equation}
Note, that variable  $t$ here is a kind of scaling parameter.

 Localization properties of the kernel $\mathcal{K}_{t}^{F}(x,y)$ are given in the following statement.

\begin{lem}\label{localiz}

 If $L$ is an elliptic self-adjoint second order differential operators on compact manifolds, then the following holds
 
 1) If $F$ is any Schwartz function on $\mathbf{R}$ , then 

\begin{equation}
\mathcal{K}_{t}^{F}(x,x)\sim c\>t^{-d},\>\>\>t\rightarrow 0.
\end{equation}

2) If, in addition, $F\in C^{\infty}_{c}(\mathbf{R})$ is even, then  on $\mathbf{\mathbf{M}} \times \mathbf{\mathbf{M}}\setminus \Delta,$ where $\Delta=\{(x,x)\},\>x\in \mathbf{\mathbf{M}}$, $\>\>\>\>\mathcal{K}_{t}^{F}(x,y)$ vanishes to infinite order as  $t$ goes to zero.

\end{lem}

 Let $g\in C^{\infty}(\mathbf{R}_{+})$ be a monotonic function such that $supp\>g\subset [0,\>  2^{2}], $ and $g(s)=1$ for $s\in [0,\>1], \>0\leq g(s)\leq 1, \>s>0.$ Setting  $G(s)=g(s)-g(2^{2}s)$ implies that $0\leq G(s)\leq 1, \>\>s\in supp\>G\subset [2^{-2},\>2^{2}].$  Clearly, $supp\>G(2^{-2j}s)\subset [2^{2j-2}, 2^{2j+2}],\>j\geq 1.$ For the functions
 $
 F_{0}(s)=\sqrt{g(s)}, \>\>F_{j}(s)=\sqrt{G(2^{-2j}s)},\>\>j\geq 1, \>\>\>
 $
 one has $\sum_{j\geq 0}F_{j}^{2}(s)=1, \>\>s\geq 0$.
 Using the spectral theorem for $L$ one  can define bounded self-adjoint operators $F_{j}(L)$ as
 $$
 F_{j}(L)f(x)=\int_{\mathbf{\mathbf{M}}}\mathcal{K}^{F}_{2^{-j}}(x,y)f(y)dy,
 $$
where 
\begin{equation}\label{kernel}
\mathcal{K}^{F}_{2^{-j}}(x,y)= \sum_{\lambda_{m}\in [2^{2j-2}, 2^{2j+2}]} F(2^{-2j}\lambda_m) u_m(x) \overline{u_m(y)}.
\end{equation}
The same spectral theorem implies  
$
\sum_{j\geq 0} F_{j}^2(L)f = f,\>\>f \in  L_{2}(\mathbf{\mathbf{M}}),
$
and taking inner product with $f$ gives
\begin{equation}
\label{norm equality-0}
\|f\|^2=\sum_{j\geq 0}\left< F_{j}^2(L)f,f\right>=\sum_{j\geq 0}\|F_{j}(L)f\|^2 .
\end{equation}
 Moreover, since the function $  F_{j}(s)$ has its support in  $
[2^{2j-2},\>\>2^{2j+2}]$ the functions $ F_{j}(L)f $ are bandlimited to  $
[2^{2j-2},\>\>2^{2j+2}]$.

Consider the sequence $\omega_{j}=2^{2j+2},\>j=0, 1, ....\>$.  
By (\ref{norm equality-0}) the equality  $
\|f\|^2=\sum_{j\geq 0}\|F_{j}(L)f\|^2 $ holds, were every  function $ F_{j}(L)f $ is bandlimited to  $
[2^{2j-2},\>\>2^{2j+2}]$.
Since for every $
\overline{F_{j}( L)f} \in \mathbf{\mathbf{E}}_{2^{2j+2}}({ L})$
one can use  Theorem \ref{product}  to conclude that
$$
|F_{j}( L)f|^2\in  \mathbf{\mathbf{E}}_{4d2^{2j+2}}({ L}),
$$
where $d=dim\>G,\>\>\mathbf{M}=G/H$.
This shows that for every $f\in L_{2}(\mathbf{M})$ we have the following decomposition
\begin{equation}
\label{addto1sc}
\sum_{j\geq \infty} \|F_{j}(L)f\|^2_2 = \|f\|^2_2,\>\>\>\>\>
|F_{j}(L)f|^2\in
\mathbf{\mathbf{E}}_{4d2^{2j+2}}({ L}).
\end{equation}
 According to Theorem \ref{cubature}  there exists a constant $a>0$ such that for all integers  $j$ if
\begin{equation}
\label{rate}
\rho_j = ad^{-1/2}2^{-j}\sim 2^{-j},\>\>d=dim\>G, \>\>\mathbf{\mathbf{M}}=G/H,
\end{equation}
then for any  $\rho_{j}$-lattice $\mathbf{\mathbf{M}}_{\rho_{j}}$ one can find coefficients $\mu_{j,k}$ with
$
\mu_{j,k}\sim \rho_j^{n},\>\>\>\>n=dim\>\mathbf{\mathbf{M}},
$
for which the following exact cubature formula holds
\begin{equation}
\label{samrate}
\|F_{j}(L)f\|^2_2 = \sum_{k=1}^{K_j}\mu_{j,k}\left|F_{j}(L)f(x_{j,k})\right|^2,
\end{equation}
where $x_{j,k} \in \mathbf{\mathbf{M}}_{\rho_j}$, $k = 1,\ldots,K_j = card\>(\mathbf{\mathbf{M}}_{\rho_j})$.
Using the kernel $\mathcal{K}_{2^{-j}}^{F}$  of the operator $F_{j}(L)$
we  define the functions
\begin{equation}
\label{vphijkdf}
\Theta_{j,k}(y) =  \sqrt{\mu_{j,k}}\>\overline{\mathcal{K}^{F}_{2^{-j}}}(x_{j,k},y) = 
$$
$$
\sqrt{\mu_{j,k}} \sum_{\lambda_{m}\in [2^{2j-2}, 2^{2j+2}]} \overline{F}(2^{-2j}\lambda_m) \overline{u}_m(x_{j,k}) u_m(y).
\end{equation}
We find that for every  $f \in L_2(\mathbf{\mathbf{M}})$ the following equality holds
$ \|f\|^2_2 = \sum_{j,k} |\langle f,\Theta_{j,k} \rangle|^2.$

\begin{thm}(Kernel localization \cite{gpes}) \label{loc}
If $\mathbf{\mathbf{M}}$ is compact  then the  functions $ \Theta_{j,k}$ are  localized around the points $x_{j,k}$ in the sense that for  any $N>0$ there exists a $C( N)>0$ such  that
  \begin{equation}\label{LOC-1}
    |\Theta_{j,k}(x)|\leq
    C( N) \frac{2^{dj}}{\max(1, \>\>2^{j}d(x, x_{j,k}))^{N}}, \end{equation}
  for all natural $ j.$
  \end{thm}
  
\begin{thm}(Bandlimited localized Parseval localized frames on homogeneous manifolds)
For any compact homogeneous manifold $\mathbf{\mathbf{M}}$ the set of functions $\{\Theta_{j,k}$\}, constructed in (\ref{vphijkdf}) forms a  Parseval frame in the Hilbert space $L_{2}(\mathbf{\mathbf{M}})$. In particular the following reconstruction formula holds true
\begin{equation}
\label{recon}
f = \sum_{j\geq 0}\sum_{k=1}^{K_{j}} \langle f,\Theta_{j,k} \rangle \Theta_{j,k},
\end{equation}
with convergence in $L_2(\mathbf{\mathbf{M}})$.  Every $\Theta_{j,k}$ is bandlimited to $[2^{2j-2}, 2^{2j+2}]$ and its localization on manifold is given by (\ref{LOC-1}).
\end{thm}

\bigskip

The condition  (\ref{rate}) imposes a specific rate of sampling in (\ref{samrate}).
It is interesting to note 
 that this rate is essentially optimal. Indeed, on one hand   the  Weyl's asymptotic formula  (\ref{Weyl}) gives the dimension of the space $\mathbf{E}_{\omega}(L)$. 
On the other hand,  the condition  (\ref{rate}) and the definition of a $\rho$-lattice imply  that 
the number of points in an "optimal" lattice $\mathbf{\mathbf{M}}_{\rho_{j}}$  for $\rho_{j}\sim 2^{-j}$ can be  approximately estimated as
$$
 card\>\mathbf{\mathbf{M}}_{\rho_{j}}\sim c\frac{Vol(\mathbf{\mathbf{M}})}{2^{-jn/2}}=cVol(\mathbf{\mathbf{M}})2^{jn/2},\>\>\>n=dim\>\mathbf{\mathbf{M}},
$$
which is in agreement with  the Weyl's formula (\ref{Weyl}) with $\omega\sim 2^{j}$.

\section{Shannon sampling of bandlimited functions}

 We consider an even $F\in C^{\infty}_{c}(\mathbf{R})$ which equals $1$ on $[-1,1]$,
and which is supported in $[-\Omega,\> \Omega],\>\>\Omega>1.$  Let $\mathcal{K}_{\Omega^{-1/2}}^{F}(x,y)$ be the kernel of $F(\Omega^{-1}L)$ defined by (\ref{kernel-1}).
 If $0<\omega\leq \Omega$ then since $F(\Omega^{-1}\lambda_k) = 1$ whenever $\lambda_k \leq  \omega$, we have that according  to (\ref{function}) - (\ref{kernel-1}) for every $f \in  \mathbf{E}_{\omega}(L)$ the
following reproducing formula holds

\begin{equation}
\label{repro}
f(x) = \left[F(\Omega^{-1}L)f\right](x) = \int_{\mathbf{M}} \mathcal{K}_{\Omega^{-1/2}}^{F}(x,y) f(y) dy 
\end{equation}
where $dy$ is the normalized invariant measure. 
Clearly, for a fixed $x\in \mathbf{M}$ the kernel $\mathcal{K}_{\Omega^{-1/2}}^{F}(x,y)$ as a function in $y$ belongs to $\mathbf{E}_{\Omega}(L)$. Thus, for $f\in \mathbf{E}_{\omega}(L),\>\>\omega<\Omega,$ the Product property (Theorem \ref{product}) implies that the product $ \mathcal{K}_{\Omega^{-1/2}}^{F}(x,y) f(y) $ belongs to $\mathbf{E}_{4d\Omega}(L)$, where $d=dim\>G$.  Now an  application of the Cubature formula (Theorem \ref{cubature}) implies the following theorem.

 \begin{thm} 
For every  compact homogeneous manifold $\mathbf{M}=G/H$ 
there exists  a constant $c=c(\mathbf{M})$ such that for any $\Omega>0$ and any lattice $\mathbf{M}_{\rho}=\{x_{k}\}_{k=1}^{m_{\Omega}}$ with $\rho=c\Omega^{-1/2}$ one can find positive weights $\mu_{k}$
$$
\mu_{k}\asymp \Omega^{-n/2},\>\>n=dim\>\mathbf{M}, 
$$
such that for any $f\in  \mathbf{E}_{\omega}(L)$ with $\omega\leq \Omega$ the following analog of the Shannon formula holds

\begin{equation}\label{Sh100} f(x) = \sum_{k = 1}^{m_{\Omega}} \mu_k f(x_k)  \mathcal{K}_{\Omega^{-1/2}}^{F}(x, x_{k}),\>\>\>f\in  \mathbf{E}_{\omega}(L).
\end{equation}
\end{thm}

\begin{rem}
Note that our definition of a $\rho$-lattice and the Weyl's asymptotic formula (\ref{Weyl}) for eigenvalues of $L$ imply  that  $m_{\Omega}$ is ``essentially" the dimension of the space 
$\mathbf{E}_{4d\Omega}(L)$ with  $d=dim\>G$. In other words there exists a constants $C_{1}(\mathbf{M})>0,\>C_{2}(\mathbf{M})>0$ (which are independent on $\Omega$) such that the number $m_{\Omega}$ of sampling points satisfies the following inequalities
\begin{equation}\label{cond-00}
  C_{1}(\mathbf{M})\Omega^{n/2}\leq m_{\Omega}\leq  C_{2}(\mathbf{M})\Omega^{n/2}
  $$
  $$
  C_{1}(\mathbf{M})\mathbf{E}_{4d\Omega}(L)\leq m_{\Omega}\leq  C_{2}(\mathbf{M})\mathbf{E}_{4d\Omega}(L).
 \end{equation}
\end{rem}
\begin{rem}
Lemma \ref{localiz} shows that for large $\Omega$ functions $\mathcal{K}_{\Omega^{-1/2}}^{F}(x, x_{k})$ in (\ref{Sh100}) are essentially localized around sampling points $x_{k}$.

\end{rem}

\section{A discrete formula for evaluating Fourier coefficients on manifolds.} 

As another application of the Product Property and the Cubature Formula, we prove an analog of the Shannon Sampling Theorem on compact homogeneous manifolds.

Theorems  \ref{product} and \ref{cubature} imply the following theorem which shows that on a compact homogeneous manifold $\mathbf{M}$ there are finite sets of points which yield exact  discrete formulas for computing Fourier coefficients of bandlimited functions. 

\begin{thm}
For every  compact homogeneous manifold $\mathbf{M}=G/H$ 
there exists  a constant $c=c(\mathbf{M})$ such that for any $\omega>0$ and any lattice $\mathbf{M}_{\rho}=\{x_{k}\}_{k=1}^{r_{\omega}}$ with $\rho=c\omega^{-1/2}$ one can find positive weights $\mu_{k}$ comparable to 
$
\omega^{-n/2},\>\>\>n=dim\>\mathbf{M},
$ 
such that Fourier coefficients $c_{i}(f)$  of any $f$ in $\mathbf{E}_{\omega}(L)$ with respect to the basis $\{u_{i}\}_{i=1}^{\infty}$ can be computed by the following \textit{exact} formula
$$
c_{i}(f)=\int_{\mathbf{M}}f(x)\overline{u_{i}}(x)dx=\sum_{k=1}^{r_{\omega}}\mu_{k}f(x_{k})\overline{u_{i}}(x_{k}),
$$
with  $r_{\omega}$ satisfying   relations 
\begin{equation}
  C_{1}(\mathbf{M})\omega^{n/2}\leq r_{\omega}\leq  C_{2}(\mathbf{M})\omega^{n/2}
  $$
  $$
  C_{1}(\mathbf{M})\mathbf{E}_{4d\omega}(L)\leq r_{\omega}\leq  C_{2}(\mathbf{M})\mathbf{E}_{4d\omega}(L),
 \end{equation}
where $C_{1}(\mathbf{M})$ and $C_{2}(\mathbf{M})$ are the same as in (\ref{cond-00}).
\end{thm}

We obviously have the following \textit{"discrete" representation formula of $f$ in $ \mathbf{\mathbf{E}}_{\omega}(L)$ in terms of eigenfunctions $u_{i}$}
\begin{equation}\label{repesentationeigenfunc}
f=\sum_{i}\sum_{k=1}^{r_{\omega}}\mu_{k}f(x_{k})\overline{u_{i}}(x_{k})u_{i}.
\end{equation}

\section*{Acknowledgment}

The work was supported in
part by the National Geospatial-Intelligence Agency University
Research Initiative (NURI), grant HM1582-08-1-0019.

\end{document}